\documentclass[11pt]{article} % use larger type; default would be 10pt

\usepackage[utf8]{inputenc} % set input encoding (not needed with XeLaTeX)

\usepackage{geometry} % to change the page dimensions \geometry{letterpaper}
\usepackage{hyperref}
\usepackage{enumerate}
\usepackage{amsmath}
\usepackage{amssymb}
\usepackage{dsfont}
\usepackage{verbatim}
\usepackage{graphicx}

\numberwithin{equation}{section}
\numberwithin{figure}{section}
\numberwithin{table}{section}

\allowdisplaybreaks

\title{Inflation Models with Correlation and Skew}
\author{\Large Orcan \"Ogetbil\footnote{orcan.ogetbil@wellsfargo.com} \ 
and Bernhard Hientzsch\footnote{bernhard.hientzsch@wellsfargo.com}\\
{\small Corporate Model Risk, Wells Fargo Bank}
}

\date{} % to remove the date from the cover

\begin{document}

\maketitle
\setcounter{secnumdepth}{3}

\begin{abstract}

We formulate a forward inflation index model with multi-factor volatility
structure featuring a parametric form that allows calibration to correlations
between indices of different tenors observed in the market. Assuming the nominal
interest rate follows a single factor Gaussian short rate model, we present
analytical prices for zero-coupon and year-on-year swaps, caps, and floors.
The same method applies to any interest rate model for which one can compute the
zero-coupon bond prices and measure shifts. We extend the multi-factor model
with leverage functions to capture the entire market volatility skew with a
single process. The time-consuming calibration step of this model can be avoided
in the simplified model that we further propose. We demonstrate the leveraged
and the simplified models with market data.

\end{abstract}

% \tableofcontents

\section{Introduction}
The general increase in the prices of goods and services in an economy is
referred as inflation. Inflation is usually measured by some consumer price
index (CPI), a weighted average of a selected set of goods and services.
As inflation has direct impact on purchasing power it constitutes an investment
risk, which can be mitigated by inflation-linked securities. The global increase
in annual inflation over the past decade has been accompanied by growth in
demand for such securities. According to \cite{Wojtowicz2023}, the global
inflation-linked bonds market size has grown by 64\% to USD 2.82 trillion
over the decade leading to April 2023.

The most common instruments traded in the inflation-linked securities market are
options on inflation index based zero-coupon and year-on-year swaps. In a
zero-coupon swap, the fixed leg payoff at the payment date is computed by
annually compounding a target rate $\bar{K}$ to maturity. The floating leg
payoff is directly proportional to the inflation index observed at the reset
date. The current US and EUR inflation markets exhibit a two to three month lag
between inflation index reset and zero-coupon swap payoff dates. In a
year-on-year swap, the floating payment depends on the ratio of two inflation
index values that are reset one year apart, and the fixed payment is based on a
simple target rate $\bar{K}_Y$.

Zero-coupon and Year-on-Year Inflation Markets might have different market
participants and could well be considered as separate segments of the market
with somewhat different factors impacting pricing in each. With sufficient data
the two markets can be modeled independently. For example, \cite{Gretarsson2013}
models the year-on-year market by a quadratic Gaussian process. As an
alternative approach \cite{MM2009} builds a framework to calibrate to both
markets in a single SABR-like model. While making a modeling choice, a
significant concern is data availability. Zero-coupon option data seems to be
more readily available than year-on-year option data. While the models we
present in this paper can be calibrated to both markets, our focus is
calibrating to zero-coupon market; in the mean time the models can price both
types of options.

In earlier work, Jarrow and Yildirim considered modeling three processes
\cite{JY2003}: real forward rate, nominal forward rate and the CPI, where the
drift of the CPI process is the difference between nominal and real interest
rates. While this formulation allows pricing of zero-coupon and year-on-year
swaps, it relies on unobservable real forward rate data for calibration. Kazziha
models inflation rates as discrete forward processes in \cite{Kazziha1999}. In
this approach, the forward inflation rate follows log-normal dynamics and is a
martingale in the associated forward measure. Log-normal dynamics have been
further analyzed in \cite{BNM2004} and \cite{Mercurio2005}. Modeling both the
short rate and the inflation index with simple Hull-White processes allows
closed form solutions \cite{DK2006}. A multi-factor volatility model with
SABR-like volatility dynamics with closed form approximations under frozen drift
assumptions were formulated by \cite{MM2009}. \cite{ACDP2019} studies the joint
evolution of interest and inflation rates as jump processes in the European
market. A Cox-Ingersoll-Ross type stochastic variance process for inflation was
introduced in \cite{MM2006}, which yields a fast Fourier transform based
analytical solution for the uncorrelated case, and an approximation with frozen
drift assumption for the correlated case. Extending real and nominal rates
modeling approach, \cite{Begot2016} proposes a stochastic local volatility
model. It however excludes details for calibration steps of leverage functions
as well as numerical demonstration.

Our first goal in this paper is to construct a multi-factor log-normal model for
inflation index that captures correlations between underliers of different
tenors observed in the market. We model the interest rate stochastically by a
one factor Gaussian process. While our approach can work with any short rate
model as long as certain quantities can be computed, here we demonstrate it with
the one factor Gaussian model G1++. We start by reviewing the Kazziha model
subject to G1++ interest rate in section \ref{sec:KazzihaG1pp}. Section
\ref{sec:instruments} introduces the zero-coupon and year-on-year swaps, caps
and floors under consideration. In section \ref{sec:multifactor}, we formulate
parametric two and three factor models which capture the market correlations,
and present closed form formulae for zero-coupon and year-on-year instruments.
Later we extend this model with leverage functions to capture the volatility
skew observed in the market. To conclude this section, we further propose a
simplified model, which avoids the calibration step required by the leveraged
model, and performs similarly well. Finally, section \ref{sec:discussion}
summarizes our findings.

\section{Kazziha inflation model under G1++ interest rate}\label{sec:KazzihaG1pp}
Before proceeding with modeling the inflation we furnish the base environment
with a short rate process by which we can compute bond prices. Let
$W^{\mathbb{Q}\text{(r)}}$ be a Brownian motion under risk-neutral measure
$\mathbb{Q}$ of filtered probability space $\left(\Omega, \mathcal{F},
\left\{\mathcal{F}_t, t \geq 0 \right\}, \mathbb{Q}\right)$. We assume that the
num\'eraire associated with the risk-neutral measure $\mathbb{Q}$, that is the
money market account $B(t)$, accrues at short rate $r(t)$ by $dB(t) = r(t) B(t)
dt$. The short rate is modeled by a Gaussian single factor process
\cite{AP2010},
\begin{equation}
\begin{split}
r_t &= x_t + \phi_t, \\
dx_t &= -a_t x_t dt + \sigma^r_t dW^{\mathbb{Q}\text{(r)}}_t.
\label{eqn:dr_G1PP_RN}
\end{split}
\end{equation}
Here $\phi_t$ is the shift function that is calibrated to market discount curve,
$a_t \geq 0$ are the mean reversion coefficients, $\sigma^r_t > 0$ are the
volatility coefficients, and $x_0 = 0$. The discount factor is given by $D(t)
\equiv 1/B(t)=\exp{\left[-\int_0^tr(u) du\right]}$.

We denote by $P(t, T) \equiv \mathbf{E}^{\mathbb{Q}} \left[\frac{D(T)}{D(t)}
\middle\vert \mathcal{F}_t \right]$ the time $t$ value of the zero coupon bond
maturing at time $T$, through which we define the instantaneous forward rate
\begin{equation}
f(t, T) \equiv - \frac{\partial \log P(t, T) }{\partial T}
= - \frac{1}{P(t, T) }\frac{\partial P(t, T) }{\partial T}.
\end{equation}
Under the $T$-forward measure $\mathbb{P}^{T}$ defined by the
num\'{e}raire $P(t, T)$, the short rate process evolves as \cite{OGH2022}
\begin{equation}
dx_t = \left[-a_t x_t - b(t, T) (\sigma^r_t)^2 \right]
dt + \sigma^r_t dW^{\mathbb{P}^T \text{(r)}}_t,\label{eqn:dr_G1PP_Tfwd}
\end{equation}
with
\begin{equation*}
b(t, T) \equiv \int_t^T e^{- \int_t^v a_z dz} dv.
\end{equation*}
Here $W^{\mathbb{P}^T \text{(r)}}$ is a Brownian motion under the $T$-forward measure
$\mathbb{P}^{T}$, and it is related to $W^{\mathbb{Q}\text{(r)}}$ by
\begin{equation}
dW^{\mathbb{P}^T \text{(r)}}_t = dW^{\mathbb{Q}\text{(r)}}_t + b(t, T) \sigma^r_t dt.
\label{eqn:dW_dT_dRN}
\end{equation}
Using It\^{o}'s lemma one can write the SDEs for the zero coupon bond and the
instantaneous forward rate in the risk neutral measure as
\begin{equation}
\begin{split}
\frac{dP(t, T)}{P(t, T)} = & r_t dt - b(t, T) \sigma^r_t dW^{\mathbb{Q}\text{(r)}}_t,\\
df(t, T) =& b(t, T) \frac{\partial b(t, T)}{\partial T} (\sigma^r_t)^2 dt + 
 \frac{\partial b(t, T)}{\partial T} \sigma^r_t dW^{\mathbb{Q}\text{(r)}}_t,
\end{split}
\end{equation}
and in the $T$-forward measure $\mathbb{P}^{T}$ as
\begin{equation}
\begin{split}
\frac{dP(t, T)}{P(t, T)} = & \left[r_t + \left(b(t, T) \sigma^r_t
 \right)^2 \right] dt - b(t, T) \sigma^r_t
 dW^{\mathbb{P}^T \text{(r)}}_t,\\
df(t, T) =& \frac{\partial b(t, T)}{\partial T} \sigma^r_t dW^{\mathbb{P}^T \text{(r)}}_t.
 \label{eqn:dP_Tfwd}
\end{split}
\end{equation}
The above zero coupon bond price is solved as
\begin{equation}
\frac{P(t, T)}{P(0, T)} = \exp{\left[\int_0^t \left(r_s
+ \frac{1}{2} \left(b(s, T) \sigma^r_s \right)^2\right) ds
- \int_0^t b(s, T) \sigma^r_s dW^{\mathbb{P}^T \text{(r)}}_s \right]},
\label{eqn:ersds}
\end{equation}
where $P(0, T)$ is the time-zero market value of the zero coupon bond
$P(t, T)$. The time $t$ value of the zero coupon bond price can be written as
\begin{equation}
P(t, T) = \exp\left[-\int_t^T\left(\phi_s - \frac{1}{2} \left(b(s, T) \sigma_s^r
\right)^2\right)ds - b(t, T) x_t\right]\label{eqn:g1pp_zcb}
\end{equation}

Given a set of time-zero discount factors $P(0, T_1), \ldots, P(0, T_N)$ at
times $T_1, \ldots, T_N$ and model parameters $a_t, \sigma_t^r$, one can compute
the shift function with
\begin{equation}
\phi_{T_n} = \frac{1}{T_n - T_{n-1}} \log
\frac{P^z(0, T_n) P(0, T_{n - 1})}{P^z(0, T_{n - 1}) P(0, T_n)},
\end{equation}
where
\begin{equation*}
P^z(t, T) = \exp \left[\frac{1}{2} \int_t^T \left(b(s, T) \sigma_s^r
\right)^2 ds - b(t, T) x_t \right].
\end{equation*}

Consider a new zero coupon bond $P(t, \bar{T})$ maturing at time $\bar{T}$, with
associated $\bar{T}$-forward measure $\mathbb{P}^{\bar{T}}$. The Radon-Nikodym
derivative reads
\begin{equation*}
\frac{d \mathbb{P}^{\bar{T}}}{d \mathbb{P}^{T}}
= \frac{P(t, T) P(0, \bar{T})}{P(t, \bar{T}) P(0, T)}.
\end{equation*}
Plugging in \eqref{eqn:ersds}, this derivative becomes
\begin{equation*}
\begin{split}
\frac{d \mathbb{P}^{\bar{T}}}{d \mathbb{P}^{T}} =
\exp \Big[ &
\int_0^t - \frac{1}{2} (\sigma^r_s)^2 \left(b^2(s, \bar{T}) - b^2(s, T) \right) ds\\
& -\int_0^t \sigma^r_s b(s, \bar{T}) dW^{\mathbb{P}^{\bar{T}} \text{(r)}}_s
+ \int_0^t \sigma^r_s b(s, T) dW^{\mathbb{P}^T \text{(r)}}_s\Big].
\end{split}
\end{equation*}
By Girsanov theorem, one sees that
\begin{equation}
dW^{\mathbb{P}^{\bar{T}} \text{(r)}}_t = dW^{\mathbb{P}^T \text{(r)}}_t +
\sigma^r_t \left(b(t, \bar{T}) - b(t, T) \right) dt
\label{eqn:TtoTbar}
\end{equation}
is a Brownian motion under $\mathbb{P}^{\bar{T}}$.

Let us denote by $I(t)$ the CPI at time $t$. Consider a single payment
fixed-float swap on the CPI. The forward CPI $F(t; T, \tilde{T})$ is defined as
the fixed amount to be set at $T$ and exchanged at time $\tilde{T}$ so that the
swap has zero value at time $t$.
\begin{equation}
F(t; T, \tilde{T}) = E^{\mathbb{P}^{\tilde{T}}}\left[I(T)
\arrowvert \mathcal{F}_t \right].
\end{equation}

We define inflation linked zero coupon bond in terms of the nominal bond
$P(t, \tilde{T})$ and the forward CPI rate $F(t; T, \tilde{T})$ as
\begin{equation*}
P_{\text{IL}}(t, \tilde{T}) \equiv P(t, \tilde{T}) \cdot F(t; T, \tilde{T}).
\end{equation*}

CPI values are announced at times $T_i = T_0, T_1, \ldots, T_I$. In Kazziha
model, the dynamics for $F_i(t) \equiv F(t; T_i, \tilde{T}_i)$ are specified
by the single-factor log-normal process \cite{Kazziha1999}
\begin{equation}
\frac{dF_i(t)}{F_i(t)} = \sigma_i dW^{\mathbb{P}^{\tilde{T}_i} \text{(F)}}_t \label{eqn:KazzihaSDE}
\end{equation} 
where $W^{\mathbb{P}^{\tilde{T}_i} \text{(F)}}$ is a Brownian motion under the
$\tilde{T}_i$-forward measure $\mathbb{P}^{\tilde{T}_i}$ with num\'eraire
$P(t, \tilde{T}_i)$.

To price derivatives involving two or more forward CPIs, one needs a common
measure. For this purpose one considers a nominal zero coupon bond $P(t, T_p)$
of maturity $T_p$. Under the measure associated with $P(t, T_p)$, the CPI
process generally has nonzero drift,
\begin{equation*}
\frac{dF_i(t)}{F_i(t)} = \mu_i(t) dt + \sigma_i dW^{\mathbb{P}^{T_p} \text{(F)}}_t.
\end{equation*}
Let $\rho_{rF}$ be the coefficient of correlation between the Brownian motions
$W^{\mathbb{P}^{T_p} \text{(r)}}$ and $W^{\mathbb{P}^{T_p} \text{(F)}}$ is
$\rho_{rF} = \frac{d}{dt}\left<W^{\mathbb{P}^{T_p} \text{(r)}},
W^{\mathbb{P}^{T_p} \text{(F)}} \right>_t$. Using \eqref{eqn:TtoTbar} and Lemma
A.1 of \cite{OGH2022}, we find that under $\mathbb{P}^{T_p}$ the CPI process
follows
\begin{equation}
\frac{dF_i(t)}{F_i(t)} = - \rho_{rF} \sigma_i \sigma^r_t \left(b(t, T_p) 
- b(t, \tilde{T}_i) \right) dt + \sigma_i dW^{\mathbb{P}^{T_p} \text{(F)}}_t,
\label{eqn:dF_Tfwd}
\end{equation}
so that $\mu_i(t) = - \rho_{rF} \sigma_i \sigma^r_t \left(b(t, T_p) 
- b(t, \tilde{T}_i) \right)$. One can use the above SDEs to compute
the expectation of the forward CPI as
\begin{equation}
E^{\mathbb{P}^{T_p}}\left[F_i({T}_i) \arrowvert \mathcal{F}_t \right]
= F_i(t) e^{\int_{t}^{{T}_i} \mu_i(s) ds}.
\end{equation}

Applying the shift in the Brownian motion \eqref{eqn:dW_dT_dRN} in the reverse
direction to the SDE \eqref{eqn:KazzihaSDE}, or setting $T_p = b(t, T_p) = 0$ in
\eqref{eqn:dF_Tfwd}, we obtain the evolution of the CPI process in risk neutral
measure
\begin{equation}
\frac{dF_i(t)}{F_i(t)} = \rho_{rF} \sigma_i \sigma^r_t b(t, \tilde{T}_i) 
 dt + \sigma_i dW^{\mathbb{Q}\text{(F)}}_t.\label{eqn:dF_RN}
\end{equation}

\section{Inflation Instruments}\label{sec:instruments}

\subsection{Zero-coupon swap, cap, floor}
Zero-coupon swaps, caps and floors are the most standard exchange traded
instruments. The general swap(let) has a single payoff at time
$\tilde{T}_i$, that depends on the inflation rate set at time $T_{i}$ as
\begin{equation*}
\text{Swap}_{i} = \bar{N} \cdot
\left[\frac{I(T_{i})}{\bar{I}} - (1 + \bar{K})^{\bar{T}} \right],
\end{equation*}
where $\bar{N}$ is the notional amount, $\bar{I}$ is the reference rate,
$\bar{K}$ is the compounded strike, and $\bar{T}$ is the tenor as contractual
quantities. Defining $N \equiv \frac{\bar{N}}{\bar{I}}$, and $K \equiv
\bar{I} \left(1+\bar{K}\right)^{\bar{T}}$, the payoff can be written as
\begin{equation}
\text{Swap}_{i}(K, \tilde{T}_i, T_i, \tilde{T}_i) = N (I(T_{i}) - K) .
\end{equation}
The time $t$ value of this swap can be evaluated analytically as
\begin{equation}
\begin{split}
\text{Swap}_i(K, t, T_i, \tilde{T}_i) =& N \cdot P(t, \tilde{T}_i) \cdot E^{\mathbb{P}^{\tilde{T}_i}}
\left[I(T_{i}) - K \middle| \mathcal{F}_t \right]\\
=& N \cdot P(t, \tilde{T}_i) \cdot \left[F_{i} (t) - K \right].
\end{split}
\end{equation}

The cap(let) and the floor(let) have a single payoff at time $\tilde{T}_i$, that
depends on the capped/floored CPI rate set at time $T_{i}$ as
\begin{equation}
\begin{split}
\text{Cap}_{i}(K, \tilde{T}_i, T_i, \tilde{T}_i) =& N \cdot \left[I(T_{i}) - K
\right]^+,\\
\text{Floor}_{i}(K, \tilde{T}_i, T_i, \tilde{T}_i) =& N \cdot \left[K - I(T_{i})
 \right]^+.\label{eqn:cap_floor_payoff}
\end{split}
\end{equation}
where the quantities are as defined for the zero coupon swap.
The time $t$ value of the cap and the floor are given by
\begin{equation}
\begin{split}
\text{Cap}_i(K, t, T_i, \tilde{T}_i) =& N \cdot P(t, \tilde{T}_i) \cdot E^{\mathbb{P}^{\tilde{T}_i}}
\left[\left(I(T_{i}) - K \right)^+ \middle| \mathcal{F}_t \right],\\
=& N \cdot P(t, \tilde{T}_i) \cdot \left[F_{i} (t)
%e^{\int_t^{T_{i}} \mu_{i}(s) ds} 
\Phi(d_1) - K \Phi(d_2) \right],\\
%& = N \cdot E^{\mathbb{Q}}
%\left[ D(t) \left( L\left(\frac{I(T_{i})}{I^{\text{ref}}} - 1 \right)
%- \left((1 + K)^T - 1 \right) \right)^+  \Big| \mathcal{F}_t \right],\\
\text{Floor}_i(K, t, T_i, \tilde{T}_i) =& N \cdot P(t, \tilde{T}_i) \cdot E^{\mathbb{P}^{\tilde{T}_i}}
\left[\left(K - I(T_{i}) \right)^+ \middle| \mathcal{F}_t \right].\\
=& N \cdot P(t, \tilde{T}_i) \cdot \left[K \Phi(-d_2)
- F_{i} (t) 
%e^{\int_t^{T_{i}} \mu_{i}(s) ds}
\Phi(-d_1) \right],\label{eqn:cap_floor_price}
%& = N \cdot E^{\mathbb{Q}}
%\left[ D(t) \left( \left((1 + K)^T - 1 \right)
%- L\left(\frac{I(T_{i})}{I^{\text{ref}}} - 1 \right)
%\right)^+  \Big| \mathcal{F}_t \right].
\end{split}
\end{equation}
where $d_1 \equiv \frac{1}{\sigma_i \sqrt{\tau_{i}}} \log \left(\frac{F_{i}(t)
%e^{\int_t^{T_{i}} \mu_{i}(s) ds}
}{K} \right)+ \frac{\sigma_i
\sqrt{\tau_{i}}}{2}$, $d_2 \equiv d_1 - \sigma_i  \sqrt{\tau_{i}}$, $\tau_{i}
\equiv T_{i} - t$, $\Phi(\cdot)$ is the cumulative Gaussian probability
distribution, the Kazziha parameter $\sigma_i$ corresponds to the market
volatility $\Sigma_{i}(K)$ for strike $K$ and maturity $T_i$, and
the zero coupon bond price $P(t, \tilde{T}_i)$ is given in \eqref{eqn:g1pp_zcb}.

\subsection{Year-on-year swap, cap, floor}
The year-on-year inflation swap(let) has a single payoff at time $T_p$, that
depends on the inflation rates set at time $T_{i}$ and $T_j$, with
$t<T_i<T_j<T_p$ as
\begin{equation}
\text{YOYSwap}_{i}(K, T_p, T_i, T_j, T_p) = N \cdot
\left[\frac{I(T_{j})}{I(T_{i})} - (1 + \bar{K}_Y) \right],
\end{equation}
where $N$ is the notional amount, and $\bar{K}_Y$ is the strike as contractual
quantities. $T_j$ is $T_i$ plus one year. The time $t$ value of this swap is
given by
\begin{equation}
\begin{split}
\text{YOYSwap}_{i}(K_Y, t, T_i, T_j, T_p) =& N\cdot P(t, T_p) \cdot E^{\mathbb{P}^{T_p}}
\left[\frac{I(T_{j})}{I(T_{i})} - K_Y \middle| \mathcal{F}_t \right]\\
=& N\cdot P(t, T_p) \cdot
\left[X_{ij}(t) - K_Y\right],
\end{split}
\end{equation}
with $K_Y \equiv 1 + \bar{K}_Y$. The expectation of the forward ratio $X_{ij}(t)
\equiv E^{\mathbb{P}^{T_p}} \left[\frac{I(T_{j})}{I(T_{i})}\middle|
\mathcal{F}_t \right]$ is calculated as
\begin{equation}
X_{ij}(t) = \frac{F_j(t)}{F_i(t)}
\exp \left[\int_t^{T_j} \sigma_j \bar{\nu}_j(s) ds - 
\int_t^{T_i} \sigma_i \bar{\nu}_i(s) ds
- \left(\sigma_i \sigma_j + \sigma_i^2 \right)
(T_i - t) \right],
\end{equation}
and
\begin{equation*}
\bar{\nu}_i(t) \equiv \rho_{rF} \sigma^r_t
\left( b(t, \tilde{T}_i) - b(t, T_p) \right).
\end{equation*}

The year-on-year cap(let) and the floor(let) have a single payoff at time $T_p$,
that depends on the capped/floored inflation rates set at times $T_{i}$ and
$T_{j}$ as
\begin{equation}
\begin{split}
\text{YOYCap}_{i}(K_Y, T_p, T_i, T_j, T_p) =& N \cdot \left[\frac{I(T_{j})}{I(T_{i})} - K_Y
\right]^+,\\
\text{YOYFloor}_{i}(K_Y, T_p, T_i, T_j, T_p) =& N \cdot \left[K_Y - \frac{I(T_{j})}{I(T_{i})}
 \right]^+,\label{eqn:yoy_cap_floor_payoff}
\end{split}
\end{equation}
where the quantities are as defined for the year-on-year swap.
The time $t$ value of the cap and the floor are computed analytically as
\begin{equation}
\begin{split}
\text{YOYCap}_i(K_Y, t, T_i, T_j, T_p) &= N \cdot P(t, T_p) \cdot E^{\mathbb{P}^{T_p}}
\left[\left(\frac{I(T_{j})}{I(T_{i})} - K_Y \right)^+ \middle| \mathcal{F}_t \right],\\
&= N \cdot P(t, T_p) \cdot \left[X_{ij} (t)
%e^{\int_t^{T_{i}} \mu_{i}(s) ds} 
\Phi(d_1) - K_Y \Phi(d_2) \right],\\
\text{YOYFloor}_i(K_Y, t, T_i, T_j, T_p) &= N \cdot P(t, T_p) \cdot E^{\mathbb{P}^{T_p}}
\left[\left(K_Y - \frac{I(T_{j})}{I(T_{i})} \right)^+ \middle| \mathcal{F}_t \right].\\
&= N \cdot P(t, T_p) \cdot \left[K_Y \Phi(-d_2)
- X_{ij} (t) 
%e^{\int_t^{T_{i}} \mu_{i}(s) ds}
\Phi(-d_1) \right],\label{eqn:yoy_cap_floor_price}
\end{split}
\end{equation}
with
\begin{equation*}
\begin{split}
d_1 \equiv & \frac{\log \left(\frac{X_{ij}(t)}{K_Y} \right)}{\sqrt{\eta_{ij}(t)}}
 + \frac{1}{2}\sqrt{\eta_{ij}(t)},\\
d_2 \equiv& d_1 - \sqrt{\eta_{ij}(t)},\\
\eta_{ij}(t) \equiv &
\sigma_j^2  (T_j - t) + 
\left(\sigma_i^2 
- 2 \sigma_i \sigma_j 
\right) (T_i - t).\\
\end{split}
\end{equation*}

\section{Multi-factor models}\label{sec:multifactor}
\subsection{Imperfect Correlations}
\begin{figure}[ht!]
    \centering \includegraphics[width=0.3283\textwidth]{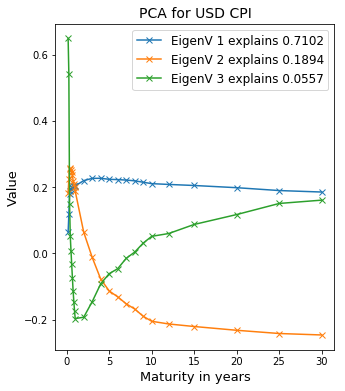}
    \includegraphics[width=0.3283\textwidth]{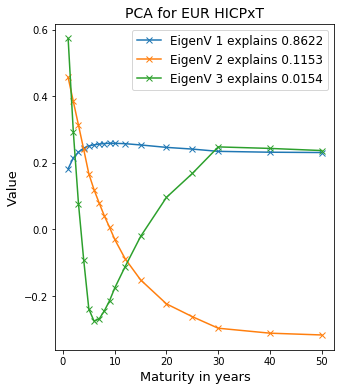}
    \includegraphics[width=0.3283\textwidth]{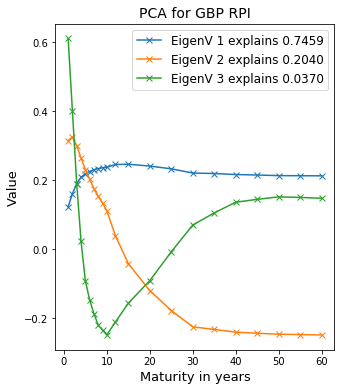} \caption[PCA
    analysis]{First three eigenvectors computed by PCA on daily returns of
    $X_k(t)$. We used historical daily closing values between 2020-03-01 and
    2023-01-23 for USD, and between 2019-11-11 and 2022-11-08 for EUR and GBP
    inflation.}
    \label{fig:pca}
\end{figure}

A significant drawback of the model \eqref{eqn:KazzihaSDE} is that it is driven
by a single Brownian factor, so that it implies perfect correlation of swap rate
returns between different maturities. We investigate the number of random
factors needed by a model to make it consistent with market correlation behavior
by doing principal component analysis (PCA) on the daily changes of swap rates,
specifically of $X_k(t) \equiv \log F_k(t)$ where the index denotes the $k$th
nearest maturity after calendar time $t$. PCA results are summarized in Figure
\ref{fig:pca}. We see that 71\%, 86\% and 75\% of the variations in curve
movements are explained by a single factor for USD, EUR, and GBP respectively.
The numbers go over 89\%, 97\%, and 94\% with two factors, and over 95\%, 99\%,
and 98\% with three factors.
% Script used:
% scripts/inflation_pca_USD.ipynb scripts/inflation_pca_EUR.ipynb
% scripts/inflation_pca_GBP.ipynb
We observe that the PCA yielded similar eigenvectors for the three inflation
curves considered. The first eigenvector is nearly constant through maturity
whereas the second and third eigenvectors contain twists that generate the
imperfect correlations. Motivated by this analysis, we decide to formulate a
model in $\tilde{T}_i$-forward measure $\mathbb{P}^{\tilde{T}_i}$ with $M$
independent random factors $W^{\mathbb{P}^{\tilde{T}_i} (F^\alpha)}, \alpha \in
\{1, \ldots, M\}$ and parameters $\mathcal{P}_M$ to incorporate imperfect market
correlations between different maturites in the inflation curve,
\begin{equation}
%\frac{dF_i(t)}{F_i(t)} = \sigma_i^1(t) dW^{\mathbb{P}^{\tilde{T}_i} (F_1)}_t
%+ \sigma_i^2(t) dW^{\mathbb{P}^{\tilde{T}_i} (F_2)}_t.
\frac{dF_i(t)}{F_i(t)} = \sigma_i \sum_{\alpha=1}^M \lambda_i^{\alpha}(t)
dW^{\mathbb{P}^{\tilde{T}_i} (F^\alpha)}_t, \label{eqn:general_multifactorSDE}
\end{equation}
with Kazziha parameter $\sigma_i$. For $M=1$ and $\lambda_i^1(t) = 1$ this
corresponds to the Kazziha model. For a two factor model setup, $M=2$, we write
\begin{equation}
\begin{split}
\lambda_i^1(t) =& 1,\\
\lambda_i^2(t) =& h_1 e^{-\kappa (T_i - t)} + h_2,\label{eqn:twofactorsSDE}
\end{split}
\end{equation}
with model parameters $\mathcal{P}_2 = \{ h_1, h_2, \kappa\}$, and $\kappa > 0$;
and for three factors, $M=3$, we extend this as
\begin{equation}
\begin{split}
\lambda_i^1(t) =& 1,\\
\lambda_i^2(t) =& h_1 e^{-\kappa_1 (T_i - t)} + h_2,\\
\lambda_i^3(t) =& h_3 (T_i - t) e^{-\kappa_2 (T_i - t)} + h_4,\label{eqn:threefactorsSDE}
\end{split}
\end{equation}
with model parameters $\mathcal{P}_3 = \{ h_1, h_2, h_3, h_4, \kappa_1,
\kappa_2 \}$, and $\kappa_1, \kappa_2 > 0$.
The multifactor model implies the following instantaneous correlation at time
$t$ between different tenors of the inflation curve,
\begin{equation}
\rho^M(t, T_i, T_j) = 
%\frac{\sum_\alpha \lambda_j^\alpha(t) \lambda_k^\alpha(t)}
%{\sqrt{\sum_\alpha \lambda_j^\alpha(t)^2} \sqrt{\sum_\alpha \lambda_k^\alpha(t)^2}}
\text{corr}\left(d\log F_i(t), d\log F_j(t)\right)
= \frac{\zeta_{ij}^M(t)}{\sqrt{\zeta^M_{ii}(t) \zeta^M_{jj}(t)}},
%= \frac{\sum_\alpha \lambda_j^\alpha(t) \lambda_k^\alpha(t)}{\zeta^M_j(t) \zeta^M_k(t)},
\end{equation}
where we defined
\begin{equation}
\zeta^M_{ij}(t) \equiv {\sum_{\alpha=1}^M \lambda_i^{\alpha}(t)\lambda_j^{\alpha}(t) }.
\end{equation}
We note that $\zeta^1_{ij}(t)=1$ for the Kazziha model. For a better fit to
historical correlation behavior, one can obtain market correlations
$\rho_{\text{market}}(T_j, T_k)$ from historical data series and then minimize
the objective function\footnote{This correlation matching methodology was
introduced in \cite{Andersen2010} for commodity underliers. Here we apply the
idea to inflation index underliers.}
\begin{equation}
J(\mathcal{P}_M ) = \sum_{j=0}^I \sum_{k=j}^I \left[\rho^M(0, T_j, T_k) -
\rho_{\text{market}}(T_j, T_k) \right]^2 .\label{eqn:objective_function}
\end{equation}
Having a set of calibrated parameters $\mathcal{P}_M$, $\sigma_i$ remains the
last parameter to determine. The total variance is computed by integrating the
log-variance of the process \eqref{eqn:general_multifactorSDE} over the lifetime
of the option. Setting the model total implied variance to the variance implied
by the market allows the model to produce market prices. In practice, one
typically sets $\sigma_i$ to match market volatilities $\Sigma_{i}$; for example
at-the-money volatilities, or volatilities corresponding to a target strike,
\begin{equation}
w^M_i(t) \equiv \Sigma_{i}^2 (T_i - t) = \sigma_i^2 \int_t^{T_i}
\zeta^M_{ii}(s) ds .\label{eqn:market_to_model_vol}
\end{equation}
The integral on the right hand side can be solved explicitly for the two-factor
model \eqref{eqn:twofactorsSDE} as
\begin{equation*}
\int_t^{T_i} \zeta^2_{ii}(s) ds = (1 + h_2^2) \tau_i
+ \frac{h_1^2}{2\kappa} \left( 1 - e^{-2\kappa \tau_i} \right)
+ \frac{2h_1 h_2}{\kappa} \left( 1 - e^{-\kappa \tau_i} \right),
\end{equation*}
and for the three-factor model \eqref{eqn:threefactorsSDE} as
\begin{equation*}
\begin{split}
\int_t^{T_i} \zeta^3_{ii}(s) ds = & (1 + h_2^2 + h_4^2) \tau_i
+ \frac{h_1^2}{2\kappa_1} \left( 1 - e^{-2\kappa_1 \tau_i} \right)
+ \frac{2h_1 h_2}{\kappa_1} \left( 1 - e^{-\kappa_1 \tau_i} \right) \\
&- \frac{h_3^2 \tau_i}{2\kappa_2^2} (\kappa_2 \tau_i + 1) e^{-2\kappa_2 \tau_i}
- \frac{2 h_3 h_4 \tau_i}{\kappa_2}  e^{-\kappa_2 \tau_i} \\
&+ \frac{h_3^2}{4\kappa_2^3} \left( 1 - e^{-2\kappa_2 \tau_i} \right)
+ \frac{2h_3 h_4}{\kappa_2^2} \left( 1 - e^{-\kappa_2 \tau_i} \right),
\end{split}
\end{equation*}
with time to maturity $\tau_i \equiv T_i - t$.

The cap and floor prices have the analytical solutions
\begin{equation}
\begin{split}
\text{Cap}_i(K, t, T_i) &= N \cdot P(t, \tilde{T}_i) \cdot \left[F_{i} (t)
\Phi(d_1) - K \Phi(d_2) \right],\\
\text{Floor}_i(K, t, T_i) &= N \cdot P(t, \tilde{T}_i) \cdot \left[K \Phi(-d_2)
- F_{i} (t)
\Phi(-d_1) \right],\label{eqn:cap_floor_price_multifactor}
\end{split}
\end{equation}
with
\begin{equation*}
\begin{split}
%\mu^M_{i} (t) \equiv& - \sigma_i \sigma^r_t \left(b(t, \tilde{T}_i) 
%- b(t, {T}_{i}) \right) \sum_{\alpha=1}^M \rho_{rF^\alpha} \lambda_i^{\alpha}(t),\\
%d_1 \equiv & \frac{1}{\sqrt{w^M_i(t)}} \left[\log \left(\frac{F_{i}(t)}{K} \right) 
%+ \int_t^{T_{i}} \mu^M_{i}(s) ds
%\right] + \frac{1}{2}\sqrt{w^M_i(t)},\\
d_1 \equiv & \frac{\log \left(\frac{F_{i}(t)}{K} \right)}{\sqrt{w^M_i(t)}}
 + \frac{1}{2}\sqrt{w^M_i(t)},\\
d_2 \equiv& d_1 - \sqrt{w^M_i(t)},\\
%w^M_i(t) \equiv& \sigma_i^2 \int_t^{T_i} \zeta^M_{ii}(s) ds.
\end{split}
\end{equation*}
%where $\rho_{rF^\alpha}$ are the coefficients of correlation between the
%Brownian motions $(W^{\mathbb{Q} (r)}_t)$ and $(W^{\mathbb{Q} (F^\alpha)}_t)$,
%that is $d\left<W^{\mathbb{Q} (r)}, W^{\mathbb{Q} (F^\alpha)} \right>_t =
%\rho_{rF^\alpha} dt$, and the zero coupon bond price $P(t, \tilde{T}_i)$ is
%given in \eqref{eqn:g1pp_zcb}.

The year-on-year cap and floor prices have the analytical solutions
\begin{equation}
\begin{split}
\text{YOYCap}_i(K_Y, t, T_i, T_j, T_p) &= N \cdot P(t, T_p) \cdot \left[X_{ij} (t)
\Phi(d_1) - K_Y \Phi(d_2) \right],\\
\text{YOYFloor}_i(K_Y, t, T_i, T_j, T_p) &= N \cdot P(t, T_p) \cdot \left[K_Y \Phi(-d_2)
- X_{ij} (t) 
\Phi(-d_1) \right],\label{eqn:yoy_cap_floor_price_multifactor}
\end{split}
\end{equation}
with
\begin{equation*}
\begin{split}
X_{ij}(t) \equiv& \frac{F_j(t)}{F_i(t)}
\exp \left[\int_t^{T_j} \sigma_j \bar{\nu}^M_j(s) ds - 
\int_t^{T_i} \left(\sigma_i \bar{\nu}^M_i(s)
- \sigma_i \sigma_j \zeta^M_{ij}(s)
- \sigma_i^2 \zeta^M_{ii}(s) \right) ds \right],\\
d_1 \equiv & \frac{\log \left(\frac{X_{ij}(t)}{K_Y} \right)}{\sqrt{\eta^M_{ij}(t)}}
 + \frac{1}{2}\sqrt{\eta^M_{ij}(t)},\\
d_2 \equiv& d_1 - \sqrt{\eta^M_{ij}(t)},\\
\bar{\nu}^M_i(t) \equiv& \sigma^r_t \left( b(t, \tilde{T}_i) - b(t, T_p) \right)
\sum_{\alpha=1}^M \rho_{rF^\alpha} \lambda_i^{\alpha}(t).\\
\end{split}
\end{equation*}
The implied variance $\eta^M_{ij}(t)$ of the year-on-year forward ratio can be
written in terms of model parameters as
\begin{equation*}
\begin{split}
\eta^M_{ij}(t) \equiv & \int_t^{T_j} 
\sigma_j^2 \zeta^M_{jj}(s) ds + 
\int_t^{T_i} \left(\sigma_i^2 \zeta^M_{ii}(s)
- 2 \sigma_i \sigma_j \zeta^M_{ij}(s) \right) ds.\\
= & w^M_j(t) + w^M_i(t)
- 2 \sigma_i \sigma_j \int_t^{T_i} \zeta^M_{ij}(s) ds.\\
\end{split}
\end{equation*}
In the ideal case broker quotes are available for options on the year-on-year
forward ratio $\frac{I(T_j)}{I(T_i)}$, one can calibrate the model parameters
$\sigma_i$ to fit the quotes. In the absence of such quotes, once can use the
$\sigma_i$s from regular cap-floors. In this case, however, the moneyness to
choose for each individual underlier $F_i(t)$ and $F_j(t)$ will have significant
impact on the year-on-year price. We will investigate this issue in subsection
\ref{sec:pricing_example}.

Before moving on we write down the evolution of the multi-factor model
\eqref{eqn:general_multifactorSDE} in the risk neutral measure $\mathbb{Q}$ as
\begin{equation}
\frac{dF_i(t)}{F_i(t)} =
\sigma_i \nu^M_i(t) dt + \sigma_i \sum_{\alpha=1}^M \lambda_i^{\alpha}(t)
dW^{\mathbb{Q} (F^\alpha)}_t,\label{eqn:dFmulti_RN}
\end{equation}
where
\begin{equation}
\nu^M_i(t) \equiv \sigma^r_t b(t, \tilde{T}_i)
\sum_{\alpha=1}^M \rho_{rF^\alpha} \lambda_i^{\alpha}(t).\label{eq:nu_multifactor}
\end{equation}

\subsubsection{Calibration Example}\label{sec:multifactor_par_calib}
\begin{figure}[ht!]
    \centering \includegraphics[width=\textwidth]{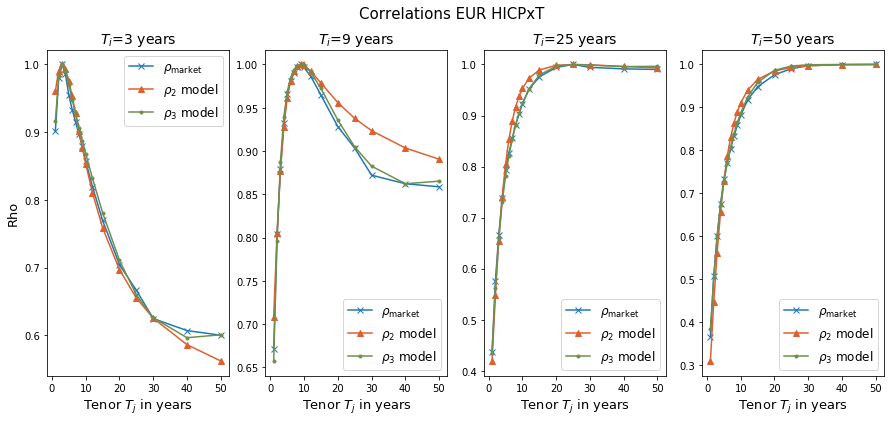}
    \caption[Model and market correlations]{Market correlations and correlations
    implied by two and three factor models for several $T_i$s as a function of
    $T_j$. Market correlations were computing using historical daily closing
    values between 2019-11-11 and 2022-11-08 for EUR inflation.}
    \label{fig:corrs-multi-factor-model}
\end{figure}

We calibrate the two and three factor models \eqref{eqn:twofactorsSDE} and
\eqref{eqn:threefactorsSDE} to historical data using scipy's \texttt{L-BFGS-B}
optimizer on the objective function \eqref{eqn:objective_function}. Figure
\ref{fig:corrs-multi-factor-model} compares the correlations implied by the two
and three factor models to historical market correlations. The two factor model
seems to capture most of the historical market correlation behavior, and
evidently the three factor model provides some additional improvement.
% Script used:
% scripts/inflation_pca_EUR.ipynb
For the two factor model the best fitting parameters are found to be
$\mathcal{P}_2 =\{ h_1, h_2, \kappa \} = \{-3.689,\  3.553,\  0.042\}$, whereas
for the three factor model they are $\mathcal{P}_3 = \{ h_1, h_2, h_3, h_4,
\kappa_1, \kappa_2 \} = \{2.319,\ -2.068,\ 0.275,\ -0.145,$ $0.085,\ 0.142\}.$

Table \ref{tbl:infmarketvol} lists the market quotes for the volatilities at
various tenors and strikes for EUR inflation index HICPxT as of 2023-04-28.
\begin{table}[ht!] \centering
\begin{tabular}{c|r|r|r|r|r|r|r|r|r}
    \multicolumn{2}{c}{} & \multicolumn{8}{|c}{$\bar{K}$}\\
    %\cline{3-10}
    \multicolumn{1}{c|}{$T_i$} & \multicolumn{1}{c|}{$F_i(0)$} & -0.02 & -0.01 & 0.00 & 0.01 & 0.02 & 0.03 & 0.04 & 0.05 \\
    \hline
    1 & 124.43 & 3.101\% & 2.756\% & 2.442\% & 2.189\% & 1.974\% & 1.839\% & 1.841\% & 1.969\% \\
    2 & 127.26 & 2.523\% & 2.242\% & 1.987\% & 1.781\% & 1.409\% & 1.293\% & 1.587\% & 1.971\% \\
    5 & 136.30 & 3.620\% & 3.218\% & 2.851\% & 2.556\% & 2.243\% & 2.415\% & 2.915\% & 3.471\% \\
    7 & 142.97 & 4.152\% & 3.691\% & 3.270\% & 2.931\% & 2.755\% & 2.986\% & 3.466\% & 4.005\% \\
    10 & 153.93 & 4.991\% & 4.437\% & 3.931\% & 3.523\% & 3.493\% & 3.789\% & 4.273\% & 4.817\% \\
    12 & 162.04 & 5.494\% & 4.884\% & 4.327\% & 3.878\% & 3.929\% & 4.253\% & 4.735\% & 5.273\% \\
    15 & 175.83 & 6.043\% & 5.371\% & 4.759\% & 4.265\% & 4.393\% & 4.735\% & 5.206\% & 5.729\% \\
    20 & 201.50 & 7.102\% & 6.313\% & 5.593\% & 5.013\% & 5.203\% & 5.548\% & 6.008\% & 6.525\% \\
\end{tabular}
\caption{Market volatility quotes $\Sigma_i(K)$ for various tenors $\bar{T}=T_i$
and strikes $\bar{K}$ with $K = F_i(0) \left(1+\bar{K}\right)^{\bar{T}}$ for
EUR inflation index HICPxT. Time is given in years.}
\label{tbl:infmarketvol}
\end{table}
We compute the volatility factors $\sigma_i$ by using at-the-money ($\bar{K}=0$)
market volatilities in \eqref{eqn:market_to_model_vol} and list them in Table
\ref{tbl:sigma_is}.
\begin{table}[ht!] \centering
\begin{tabular}{c|r|r|r}
& \multicolumn{3}{|c}{$\sigma_i$}\\
$T_i$ (years)& $M=1$ & $M=2$ & $M=3$\\
\hline
1&0.02925&0.02916&0.02404\\
2&0.02178&0.02170&0.01952\\
5&0.02961&0.02836&0.02595\\
7&0.03360&0.03070&0.02795\\
10&0.04007&0.03363&0.03091\\
12&0.04396&0.03477&0.03245\\
15&0.04820&0.03496&0.03357\\
20&0.05647&0.03598&0.03634\\
\end{tabular}
\caption{Inflation volatility factors $\sigma_i$ calibrated for the 1, 2 and 3
factor models \eqref{eqn:general_multifactorSDE}}
\label{tbl:sigma_is}
\end{table}

\subsection{Leverage functions}

The multi-factor model we introduced in the last subsection aims to capture
cross-tenor correlations as well as market volatilities for a target strike. In
order to capture the market volatility smile, that is to reprice market quotes
of different strikes, we extend the model \eqref{eqn:general_multifactorSDE}
with unique leverage functions $L_i$ for each tenor,
\begin{equation}
\frac{dF_i(t)}{F_i(t)} = L_i(F_i(t), t) \sum_{\alpha=1}^M
\lambda_i^{\alpha}(t) dW^{\mathbb{P}^{\tilde{T}_i} (F^\alpha)}_t.
\label{eqn:general_leverage_multifactorSDE}
\end{equation}
This model evolves in the risk neutral measure $\mathbb{Q}$ as
\begin{equation}
\frac{dF_i(t)}{F_i(t)} = \nu^M_i(t) L_i(F_i(t), t) dt + L_i(F_i(t), t)
\sum_{\alpha=1}^M \lambda_i^{\alpha}(t) dW^{\mathbb{Q} (F^\alpha)}_t,
\label{eqn:general_leverage_multifactorSDE_RN}
\end{equation}
where $\nu^M_i(t)$ is defined as in \eqref{eq:nu_multifactor}. The leverage
functions are to be calibrated to market quotes. They are related to time-zero
prices of $T$-maturity cap(let)s paying \eqref{eqn:cap_floor_payoff} at time
$\tilde{T}_i \geq T$ by the Dupire equation \cite{OH2023},
\begin{equation}
L_i(K, T)^2 = \frac{\frac{\partial \text{Cap}_i (K, 0, T, \tilde{T}_i)}{\partial T}
+ \theta_i^{\text{Cap}}(K, T)
}{\frac{1}{2}K^2\frac{\partial^2 \text{Cap}_i (K, 0, T, \tilde{T}_i)}{\partial K^2}
\zeta^M_{ii}(T)}, \label{eqn:dupire_caplet}
\end{equation}
where
\begin{equation*}
\begin{split}
\theta_i^{\text{Cap}}(K, T) \equiv& N \mathbf{E}^{\mathbb{Q}}\left[ D(\tilde{T}_i)
\left\{\left[F_i(T) - K\right] r_T - \nu^M_i(T)
L_i(F_i(T), T) F_i(T) \right\} \mathds{1}_{F_i(T) > K} \right]\\
& - f(0, T) \text{Cap}_i (K, 0, T, \tilde{T}_i) .\label{eqn:theta_cap}
\end{split}
\end{equation*}
In terms of floor(let)s with payoff \eqref{eqn:cap_floor_payoff}, the leverage
functions are,
\begin{equation}
L_i(K, T)^2 = \frac{\frac{\partial \text{Floor}_i (K, 0, T, \tilde{T}_i)}{\partial T}
+ \theta_i^{\text{Floor}}(K, T)
}{\frac{1}{2}K^2\frac{\partial^2 \text{Floor}_i (K, 0, T, \tilde{T}_i)}{\partial K^2}
\zeta^M_{ii}(T)
},\label{eqn:dupire_floorlet}
\end{equation}
where
\begin{equation*}
\begin{split}
\theta_i^{\text{Floor}}(K, T) \equiv& N
\mathbf{E}^{\mathbb{Q}}\left[ D(\tilde{T}_i) \left\{\left[ K - F_i(T)\right]
r_T + \nu^M_i(T) L_i(F_i(T), T) F_i(T)
\right\} \mathds{1}_{F_i(T) < K} \right]\\
& - f(0, T) \text{Floor}_i (K, 0, T, \tilde{T}_i) .\label{eqn:theta_floor}
\end{split}
\end{equation*}

The time-zero price function for a time-$T$ maturity caplet with underlier $F_i$
that pays at time $\tilde{T}_i$ can be parametrized in terms of log-moneyness
$y = \log\frac{K}{F_i(0)}$ and the total implied variance $w_i$ as
\cite{Gatheral2006}
\begin{equation}
\text{Cap}_i (y, w_i) = N P(0, \tilde{T}_i) F_i(0) \left[\Phi(d_1) - e^y \Phi(d_2)\right]
\end{equation}
where $d_1 = -y w_i^{-\frac{1}{2}} + \frac{1}{2} w_i^{\frac{1}{2}}$, and
$d_2 = d_1 - w_i^{\frac{1}{2}}$.
In the total implied variance parametrization, the Dupire equation
\ref{eqn:dupire_caplet} can be casted to $\bar{L}_i(y, T) = L_i(F_i(0) e^y, T)$
as \cite{OH2023},
\begin{equation}
\bar{L}_i(y, T)^2 = \frac{\frac{\partial \text{Cap}_i}{\partial w_i}
\frac{\partial w_i}{\partial T}
+ \theta^{\text{Cap}}_i(F_i(0) e^y, T)
}{\frac{\partial \text{Cap}_i}{\partial w_i} \left[
1 - \frac{y}{w_i} \frac{\partial w_i}{\partial y} + \frac{1}{2}
\frac{\partial^2 w_i}{\partial y^2}
+\frac{1}{4} \left(\frac{\partial w_i}{\partial y}\right)^2
\left(-\frac{1}{4}- \frac{1}{w_i} + \frac{y^2}{w_i^2}\right)
\right] \zeta^M_{ii}(T)
},
\label{eqn:dupire_caplet_tiv}
\end{equation}
with
\begin{equation*}
\frac{\partial \text{Cap}_i}{\partial w_i} = \frac{1}{2} N
P(0, \tilde{T}_i) F_i(0) e^y \Phi'(d_2) w_i^{-\frac{1}{2}}.
\end{equation*}
Similary, the time-zero price function for a time-$T$ maturity floorlet
underlier $F_i$ that pays at time $\tilde{T}_i$ can be parametrized as
\begin{equation}
\text{Floor}_i (y, w_i) = N P(0, \tilde{T}_i) F_i(0) \left[-\Phi(-d_1)
+ e^y \Phi(-d_2)\right].
\end{equation}
In the total implied variance parametrization, the Dupire equation becomes
\begin{equation}
\bar{L}_i(y, T)^2 = \frac{\frac{\partial \text{Floor}_i}{\partial w_i}
\frac{\partial w_i}{\partial T}
+ \theta^{\text{Floor}}_i(F_i(0) e^y, T)
}{\frac{\partial \text{Floor}_i}{\partial w_i} \left[
1 - \frac{y}{w_i} \frac{\partial w_i}{\partial y} + \frac{1}{2}
\frac{\partial^2 w_i}{\partial y^2}
+\frac{1}{4} \left(\frac{\partial w_i}{\partial y}\right)^2
\left(-\frac{1}{4}- \frac{1}{w_i} + \frac{y^2}{w_i^2}\right)
\right] \zeta^M_{ii}(T)
},
\label{eqn:dupire_floorlet_tiv}
\end{equation}
with
\begin{equation*}
\frac{\partial \text{Floor}_i}{\partial w_i} = \frac{1}{2} N
P(0, \tilde{T}_i) F_i(0) e^y \Phi'(-d_2) w_i^{-\frac{1}{2}}.
\end{equation*}

In the limit the interest rate volatility $\sigma^r_t$ approaches zero one has
$\nu^M_i(t) = \theta^{\text{Cap}}_i = \theta^{\text{Floor}}_i = 0$, and the
expression for the leverage function simplifies to
\begin{equation}
\bar{L}^{s}_i(y, T){}^2 = \frac{\frac{\partial w_i}{\partial T}}
{ \left[
1 - \frac{y}{w_i} \frac{\partial w_i}{\partial y} + \frac{1}{2}
\frac{\partial^2 w_i}{\partial y^2}
+\frac{1}{4} \left(\frac{\partial w_i}{\partial y}\right)^2
\left(-\frac{1}{4}- \frac{1}{w_i} + \frac{y^2}{w_i^2}\right)
\right] \zeta^M_{ii}(T)
}.
\label{eqn:dupire_caplet_deterministic_tiv}
\end{equation}
We use this expression in the calibration routine below for computing
the leverage function at time $t$ close to initial time, where the
interest rate is observed at a fixed value.

Inflation options traded on the market written on $F_i$ typically have a single
maturity $T_i$. Accordingly, the market implied volatility $\Sigma_i(K)$ for
strike $K$ yields a total implied variance $\Sigma_i(K)^2 T_i$ at maturity
$T_i$. Here we make the assumption that the total implied variance accumulates
linearly in time as $w_i = \Sigma_i(K)^2 T$ for times $T \leq T_i$,
\begin{equation}
w_i(y, T) = \Sigma_i(F_i(0) e^y)^2 T. \label{eq:tiv_def}
\end{equation}

We adapt the calibration approach proposed in \cite{OGH2022} to compute the
leverage functions $\bar{L}_i$ for every underlier $F_i$ simultaneously time
slice by time slice. We perform a Monte Carlo simulation to estimate the
expectation appearing in the expressions for $\theta_i^{\text{Cap}}(K, T)$ and
$\theta_i^{\text{Floor}}(K, T)$.

\paragraph{Inputs for calibration}
Our calibration routine expects the following quantities as input for leverage
function calibration:
\begin{itemize}
\item A multi-factor model \eqref{eqn:general_multifactorSDE} with parameters
calibrated to market data
\item Market CPI rates $F_i(0)$ as of the valuation time $t=0$
\item Market implied volatility $\Sigma_i(K)$ for each maturity
\item Market yield curve $P(0, T)$
\item G1++ short rate model \eqref{eqn:dr_G1PP_RN} (or any rate model for which
one can compute the zero-coupon bond price and the measure shift terms) with
parameters calibrated to market data.
\item Coefficients of correlation between the Brownian motions of the short rate
and the multi-factor inflation models
\end{itemize}

\paragraph{Steps for calibration}

We calibrate the leverage functions time slice by time slice, in a bootstrapping
fashion. Let $t_k; k=1, \ldots, n$ be the increasing sequence of (positive)
times where we will perform the calibration.
\begin{enumerate}
  \item Using the market implied volatilities $\Sigma_i(K) = \Sigma_i(F_i(0)
  e^y)$, generate a total implied variance surface $w_i(y, T)$ interpolator. The
  interpolator must be able to compute the partial derivatives appearing in the
  local volatility expressions.
  \item For the first time slice $t_1$, evaluate the simplified equation
  \eqref{eqn:dupire_caplet_deterministic_tiv} to compute the leverage function
  values $\bar{L}_i(y, t_1)$ for a predetermined range of strikes for every $i$.
  This step requires no Monte Carlo simulation. As a result, obtain leverage
  function values to be used until time $t_2$ in the subsequent calibration
  steps.
  \item For each of the subsequent time slices $t_k, k > 1$,  Simulate the SDE
  system \eqref{eqn:general_multifactorSDE}, \eqref{eqn:dr_G1PP_RN} up to time
  $t_k$. By choosing out-of-money options for a predetermined grid of strikes,
  that is caps for $y>0$ and floors for $y<0$, compute the Monte Carlo
  estimate for the expectation appearing in \eqref{eqn:theta_cap} or
  \eqref{eqn:theta_floor} for every $i$. Obtain the leverage function values
  $\bar{L}_i(y, t_k)$ from these equations. These values will be used during
  subsequent simulation steps from time $t_k$ to time $t_{k + 1}$. This step is
  first performed with $k=2$ and is then repeated for the remaining time slices.
\end{enumerate}

\subsubsection{Calibration Example}
We calibrate the multifactor model to market data as of 2023-04-28. We use the
same multifactor model parameters we estimated in subsection
\ref{sec:multifactor_par_calib}, and the same volatility data from Table
\ref{tbl:infmarketvol}. For simplicity we ignore the market lag, as is common in
the examples in the literature, such that $T_i = \tilde{T}_i\ \forall i$. The
G1++ model parameters are fit to market interest rate swaptions. Here we do not
go into details of this fitting. Instead we list the parameters that we use as
input, and refer to \cite{GurrieriNakabayashiWong2009} for calibration of
Hull-White-type models with time-dependent parameters. G1++ mean reversion is
set to be constant, $a_t = 0.02$. The market discount curve and G1++ volatility
parameters are given in Table \ref{tbl:marketdfg1pp}.
\begin{table}[ht!] \centering
\begin{tabular}[t]{c|l}
    \multicolumn{1}{c|}{$T$ (years)} & \multicolumn{1}{c}{$P(0, T)$} \\
    \hline
    0 & 1 \\
    1 & 0.9656 \\
    2 & 0.9379 \\
    5 & 0.8706 \\
    7 & 0.8264 \\
    10 & 0.7596 \\
    12 & 0.7152 \\
    15 & 0.6547 \\
    20 & 0.5800 \\
\end{tabular}
\quad
\begin{tabular}[t]{c|r}
    \multicolumn{1}{c|}{$t$ (years)} & \multicolumn{1}{c}{$\sigma^r_t$} \\
    \hline
    1 & 1.071\% \\
    2 & 1.093\% \\
    3 & 0.992\% \\
    5 & 0.839\% \\
    10 & 0.686\% \\
    20 & 0.683\% \\
    %20 & 0.683\% \\
\end{tabular}
\caption{Discount factors $P(0, T)$ and G1++ model volatility $\sigma^r_t$. }
\label{tbl:marketdfg1pp}
\end{table}
The coefficient of correlation between the Brownian motions $W^{\mathbb{Q}
(r)}$ and $W^{\mathbb{Q} (F^\alpha)}$ is $\rho_{rF^\alpha} =
\frac{d}{dt}\left<W^{\mathbb{Q} (r)}, W^{\mathbb{Q} (F^\alpha)} \right>_t =
-0.5\ \forall \alpha$.

The leverage functions strike grid is chosen to cover regions of concern. In our
implementation, we construct a uniform grid for $\bar{K}$ between -0.02 and 0.05
with spacing 0.001. This is translated to log-moneyness as $y = \bar{T} \log(1 +
\bar{K})$ for contractual maturity $\bar{T}$. It is important that the chosen grid is
covered by the implied volatility data. For the maturity coordinate we first
construct a time grid with uniform spacing, e.g. $t_{k+1} - t_k = 1/4$ until the
latest maturity, and then we add the quoted option maturity times to this grid.
The expectation in the leverage equation is estimated by simulating 2000
paths and computing Monte Carlo averages of the argument of the
expectation.

\begin{figure}[ht!]
    \centering
    \includegraphics[width=\textwidth]{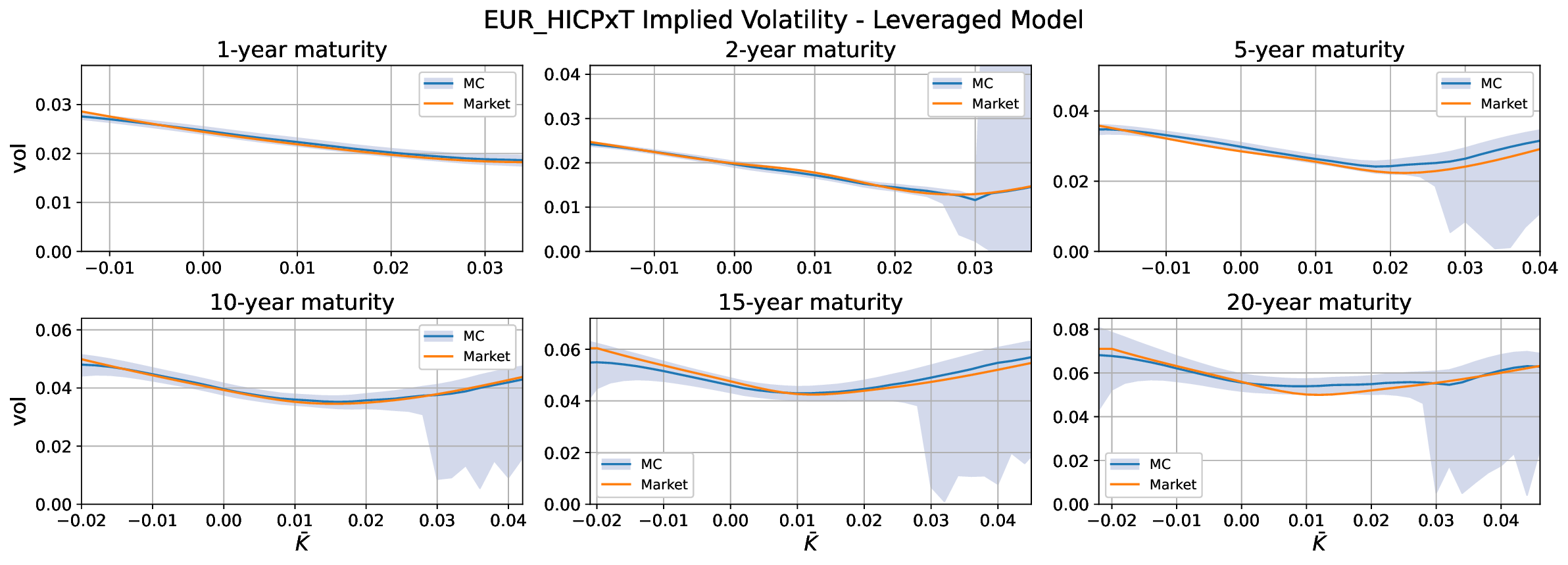}
    \caption[Implied vol recovery]{Market and Monte Carlo implied volatilities
    for the leveraged model \eqref{eqn:general_leverage_multifactorSDE} with
    $M=3$ factors \eqref{eqn:threefactorsSDE} for EUR inflation as of
    2023-04-28. The market implied volatilities are within two Monte Carlo
    standard errors (blue shaded regions) for most strikes.
    %The differences
    %in the far wings, where the price sensitivity to volatility is low, are due
    %to the near degeneracy of the pricing formula inversion.
    }
    \label{fig:implied_vol_recovery_localvol}
\end{figure}

We simulate the calibrated model over 2000 paths to price caps at various
maturities and strikes. The leverage function values are interpolated piecewise
linearly in both dimensions during simulation. We invert the pricing formula
\eqref{eqn:cap_floor_price_multifactor} to compute the model implied
volatilities from the Monte Carlo price means, as well as price means bumped by
two Monte Carlo standard errors in both directions. Figure
\ref{fig:implied_vol_recovery_localvol} shows that the market implied
volatilities are within two Monte Carlo errors of the simulation means for most
strikes within the test range.
% Script used:
% scripts/localvol_g1pp_EUR.ipynb
This test demonstrates that the implementation of the leveraged three factor
model recovers market quotes at various strikes and maturities.

\subsection{Simplified Model}
The leveraged model of the previous subsection seems to capture the market skew
for caps and floors well. The calibration routine of the leverage function, however,
involves a Monte Carlo estimation. Here we formulate a simplified model by ignoring
negligible terms in the Dupire equation such that the resulting model does not
require the calibration step. 

We can approximate the leverage function by
\begin{equation}
\bar{L}_i(y, T)^2 \approx \Lambda_i(y, T)^2 \equiv \frac{\frac{\partial
w_i}{\partial T}} {\left[1 - \frac{1}{2} \frac{y}{w_i} \frac{\partial
w_i}{\partial y} \right]^2 \zeta^M_{ii}(T)}.\label{eq:lambda_1}
\end{equation}
By plugging in the expression \eqref{eq:tiv_def} for total implied variance,
\eqref{eq:lambda_1} can be written as
\begin{equation}
\Lambda_i(y, T) = \frac{q_i(F_i(0) e^y)}{\sqrt{\zeta^M_{ii}(T)}},
\end{equation}
where
\begin{equation}
q_i(K) \equiv
\frac{\Sigma_i(K)}{1 - \frac{K\log{\frac{K}{F_i(0)}}}{\Sigma_i(K)}
\frac{\partial \Sigma_i}{\partial K}}.
\end{equation}
%In this parametrization $\Sigma_i(K)$ is the implied volatility for maturity
%$T_i$ and strike $K$.
With this function, we can formulate a simplified multi-factor model as
\begin{equation}
\frac{dF_i(t)}{F_i(t)}
=\frac{q_i(F_i(t))}{\sqrt{\zeta^M_{ii}(t)}}
\sum_{\alpha=1}^M \lambda_i^{\alpha}(t)
dW^{\mathbb{P}^{\tilde{T}_i} (F^\alpha)}_t.\label{eqn:sde_simplified}
\end{equation}
%We note that the volatility factors $\sigma_i$ of
%\eqref{eqn:general_leverage_multifactorSDE} cancel out in the final SDE of the
%simplified model.

In practice we use an algorithmic cap parameter $\eta$ for
$q_i(K)$,
\begin{equation}
q_i(K) \equiv \frac{\Sigma_i(K)}{\max\left(\frac{1}{\eta}, 1 -
\frac{K\log{\frac{K}{F_i(0)}}}{\Sigma_i(K)}
\frac{\partial \Sigma_i}{\partial K} \right)}.
\end{equation}
Our testing and analysis shows that $\eta=10$ is typically a good choice.

\begin{figure}[ht!]
    \centering
    \includegraphics[width=\textwidth]{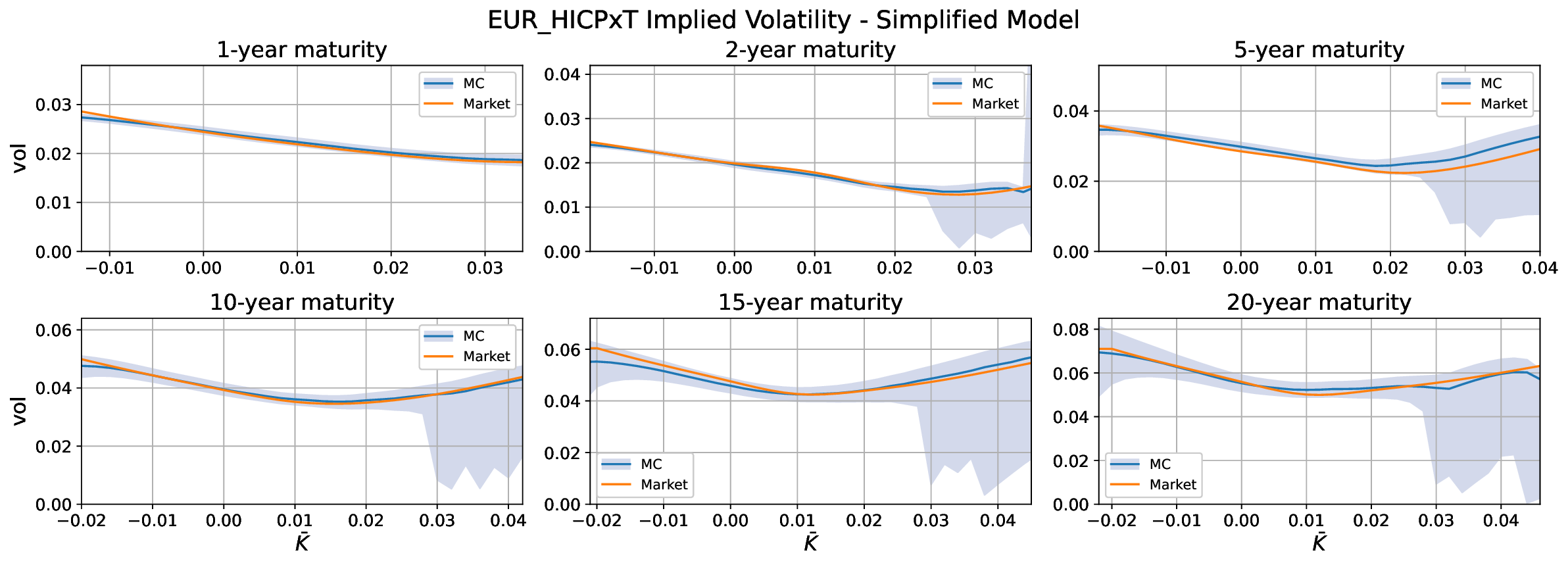}
    \caption[Implied vol recovery]{Market and Monte Carlo implied volatilities
    for the simplified model \eqref{eqn:sde_simplified} with $M=3$ factors
    \eqref{eqn:threefactorsSDE} for EUR inflation as of 2023-04-28. The market
    implied volatilities are within two Monte Carlo standard errors (blue shaded
    regions) for most strikes.}
    \label{fig:implied_vol_recovery_lvi}
\end{figure}

As in the previous subsection, we simulate the calibrated model over 2000 paths
to price caps at various maturities and strikes. We compute the model implied
volatilities from the Monte Carlo prices by inverting the pricing formula
\eqref{eqn:cap_floor_price_multifactor}. Figure
\ref{fig:implied_vol_recovery_lvi} shows that the market implied volatilities
are within two Monte Carlo errors of the simulation means for most strikes
within the test range. Moreover comparison to Figure
\ref{fig:implied_vol_recovery_localvol} reveals that the simplified model
performs similarly to the leveraged model in terms of accuracy.
% Script used:
% scripts/lvi_g1pp_EUR.ipynb

\subsection{Pricing Example}\label{sec:pricing_example}

We simulate the leveraged model \eqref{eqn:general_leverage_multifactorSDE} and
the simplified model \eqref{eqn:sde_simplified} with 3 factors over 2000 paths
to price 1-year to 2-year caps for several strikes with payoff defined in
\eqref{eqn:yoy_cap_floor_payoff}. We compare the Monte Carlo prices to the
analytical prices given by \eqref{eqn:yoy_cap_floor_price_multifactor}.
We note that the strike $K_Y$ of the year-on-year contract does not directly
correspond to the strike $K$ that goes in $\Sigma_i(K)$ while calibrating the
Kazziha parameter $\sigma_i$ to market volatilities by
\eqref{eqn:market_to_model_vol} for underlier $F_i$. If regular cap/floor
quotes, e.g. $\Sigma_i(K)$, are only what is available as market data, one needs
to pick a moneyness for the underlier $F_i$ to compute $\sigma_i$. Here we study
the impact of this choice by computing analytical prices with $\bar{K}$ ranging
from -0.02 to 0.03, where $K = F_i(0) (1 + \bar{K})^{T_i}$.

\begin{figure}[ht!]
    \centering
    \includegraphics[width=\textwidth]{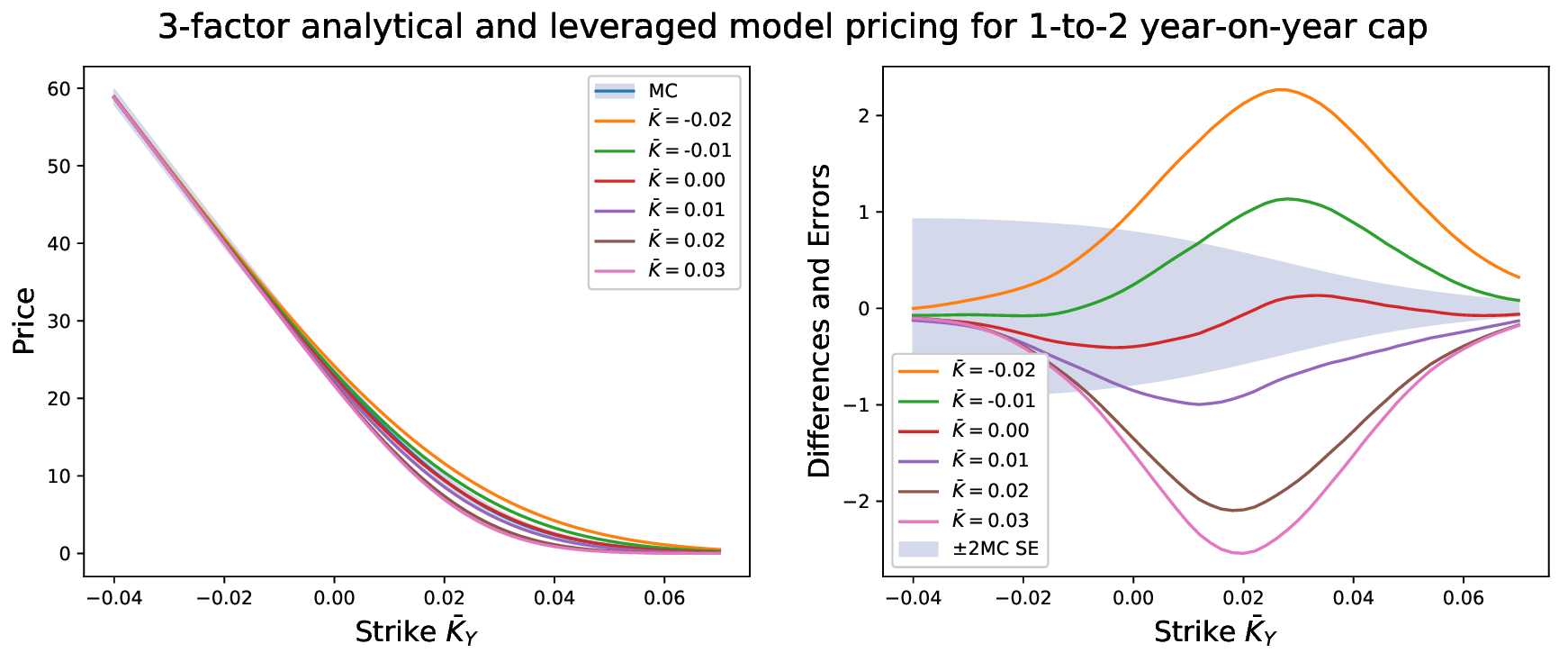}
    \caption[YoY cap pricing]{Monte Carlo 1-year to 2-year cap prices for the
    leveraged model \eqref{eqn:general_leverage_multifactorSDE} with $M=3$
    factors \eqref{eqn:threefactorsSDE} for EUR inflation as of 2023-04-28 and
    notional $N=1000$, compared to analytical prices computed with several
    individual underlier moneynesses $\bar{K}$. The analytical prices are within
    two Monte Carlo standard errors (blue shaded regions) for most strikes when
    individual underlier moneynesses are chosen near at-the-money
    ($\bar{K}\rightarrow 0$).}
    \label{fig:yoy_pricing_lv}
\end{figure}

\begin{figure}[ht!]
    \centering
    \includegraphics[width=\textwidth]{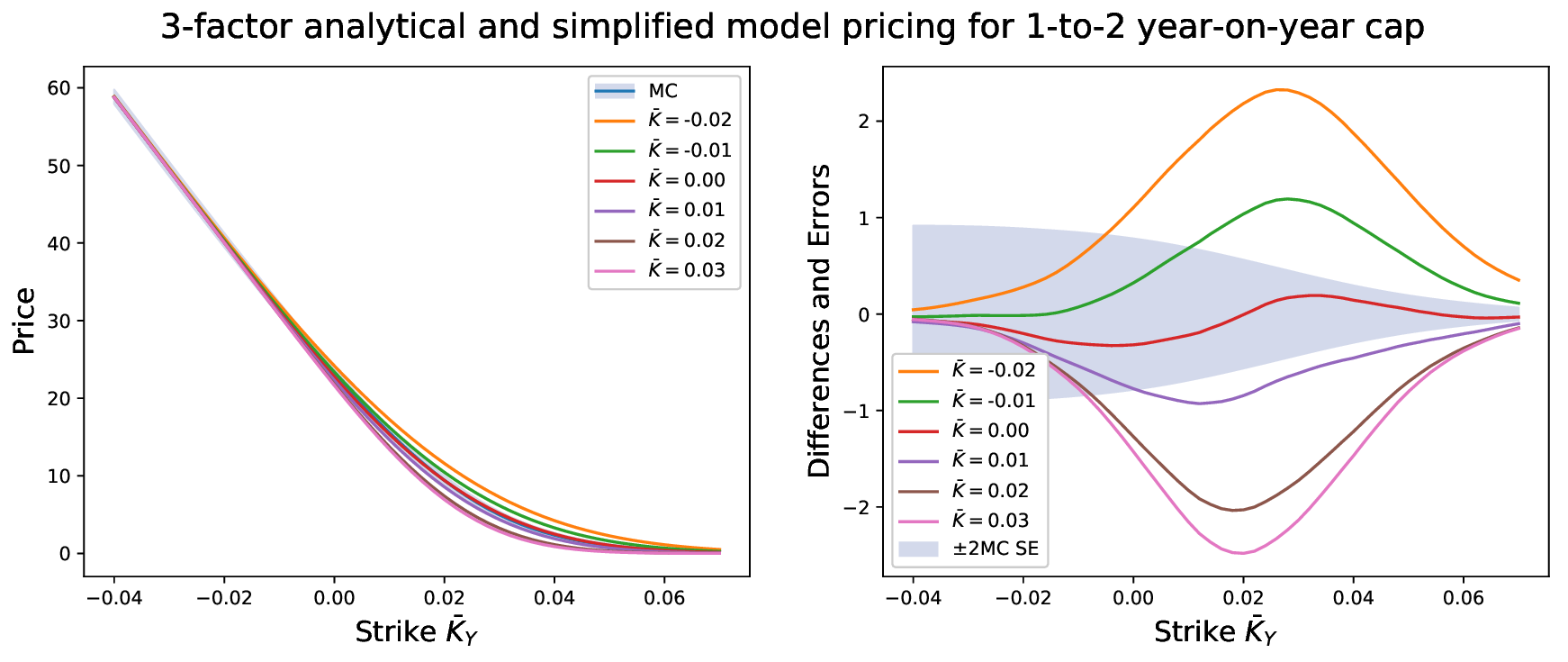}
    \caption[YoY cap pricing]{Monte Carlo 1-year to 2-year cap prices for the
    simplified model \eqref{eqn:sde_simplified} with $M=3$ factors
    \eqref{eqn:threefactorsSDE} for EUR inflation as of 2023-04-28 and notional
    $N=1000$, compared to analytical prices computed with several individual
    underlier moneynesses. The analytical prices are within two Monte Carlo
    standard errors (blue shaded regions) for most strikes when individual
    underlier moneynesses are chosen near at-the-money ($\bar{K}\rightarrow 0$).}
    \label{fig:yoy_pricing_lvi}
\end{figure}

Figure \ref{fig:yoy_pricing_lv} compares the Monte Carlo prices for the
leveraged model to the analytical prices computed by
\eqref{eqn:yoy_cap_floor_price_multifactor}. Similarly, Figure
\ref{fig:yoy_pricing_lvi} shows the comparison between the Monte Carlo prices
for the simplified model to the analytical prices. The first observation is that
both the leveraged and the simplified model give similar prices. The second
observation is, the analytical model prices vary significantly by the choice of
individual underlier moneynesses ($\bar{K}$) when calibrating the Kazziha
parameter $\sigma_i$ to market volatilities $\Sigma_i(K)$. For both the
leveraged and the simplified models, the simulated model prices are close to the
analytical prices, that is the differences are within two standard errors for
most strikes in the test range, only if we choose the individual underlier
moneynesses close to at-the-money ($\bar{K}\rightarrow 0$) during the analytical
price computation.

% Script used:
% scripts/localvol_g1pp_EUR_yoy.ipynb

% Script used:
% scripts/lvi_g1pp_EUR_yoy.ipynb

\section{Discussion}\label{sec:discussion}
After reviewing the dynamics of the single-factor Kazziha inflation model
coupled with single-factor G1++ interest rate model, we explored multi-factor
inflation models that can be calibrated to historical correlations between
inflation index underliers of available maturities. The multi-factor model
\eqref{eqn:general_multifactorSDE} that we propose is analytically tractable,
and we give closed form solutions for the prices of the inflation instruments
under consideration. This model can be calibrated to regular or year-on-year
caps/floors provided there is market data available. If year-on-year market data
is unavailable, one can still calibrate individual underliers to available
regular cap/floor market data. In this case, however, the individual underlier
moneyness that one picks has an impact on the analytical year-on-year cap/floor
price.

We further extended the multi-factor model with leverage functions
\eqref{eqn:general_leverage_multifactorSDE} to capture the market skew
nonparametrically. We proposed a slice-by-slice calibration methodology for the
leverage functions using Monte Carlo simulation, and demonstrated that the
calibrated model captures the market skew in volatility. By ignoring terms of
small size in the expression of the leverage function, we formulated a
simplified multi-factor model \eqref{eqn:sde_simplified} which does not require
the calibration step. Remarkably the simplified model replicates the market skew
as well as the full leveraged model, and also produces similar year-on-year
cap/floor prices. We compared the year-on-year cap prices of the leveraged and
simplified models to the analytical prices with Kazziha parameter calibrated to
several underlier moneynesses. We observed that the analytical prices are closer
to leveraged or simplified model prices when the individual underlier
moneynesses are chosen close to zero. The leveraged and simplified models do not
have this ambiguity as they are constructed to capture the skew at all
moneynesses, instead of a target moneyness.

\paragraph{Acknowledgments}
The authors thank Agus Sudjianto for his support on research throughout the
years leading into this work, and Vijayan Nair for his comments and suggestions
regarding this work. Any opinions, findings and conclusions or recommendations
expressed in this material are those of the authors and do not necessarily
reflect the views of Wells Fargo Bank, N.A., its parent company, affiliates and
subsidiaries.

\bibliographystyle{unsrt}
\bibliography{inflation}

\end{document}